
\magnification=1250
\baselineskip=20pt
\overfullrule=0pt

\centerline{\bf SEMICLASSICAL ANALYSIS OF TWO- AND THREE-SPIN}
\centerline{\bf ANTIFERROMAGNETS AND ANYONS ON A SPHERE}

\vskip .4in

\centerline{DIPTIMAN ~SEN\footnote{*}{E-mail: ~diptiman@cts.iisc.ernet.in}}
\vskip .2in

\centerline{\it Centre for Theoretical Studies, Indian Institute of Science,}
\centerline{\it Bangalore 560012, India}

\vskip .4in

\def\sc{semiclassical~}
\def\qu{quantum~}
\def\ha{Hamiltonian~}
\def\la{Lagrangian~}
\def\wz{Wess-Zumino~}
\def\wf{wave functions~}
\def\pain{path integral~}
\def\ex{exchange~}
\def\bg{braid group~}
\def\pg{permutation group~}
\def\bn{$~B_N~$}
\def\sn{$~S_N~$}
\def\son{{\vec S_1}}
\def\stw{{\vec S_2}}
\def\sth{{\vec S_3}}
\def\lo{{\vec l}}
\def\ph{{\vec \phi}}
\def\phon{{\vec \phi_1}}
\def\phtw{{\vec \phi_2}}
\def\phth{{\vec \phi_3}}
\def\romat{{\underline R}}
\def\vmat{{\underline V}}

\line{\bf Abstract \hfill}

\vskip .3in

We do a semiclassical analysis for two or three spins which
are coupled antiferromagnetically to each other. The semiclassical wave
functions transform correctly under
permutations of the spins if one takes into account the Wess-Zumino term
present in the path integral for spins. The Wess-Zumino term here is
a total derivative which has no effect on the energy spectrum.
The semiclassical problem is related to that of anyons moving on a sphere
with the statistics parameter $\theta$ being $2 \pi S$
for two spins and $3 \pi S$ for three spins. Finally, we present a novel way
of deriving the semiclassical wave functions from the spin wave functions.

\vskip .5in


\vfill
\eject

This Letter illustrates three different ideas using a simple \qu
mechanical model. These ideas, which have been studied recently in other
contexts, include a \sc treatment of \qu
spins for large values of the spin $S$ [~1~-~5~], a \wz (WZ) term which is
present in the \pain for spins [~2,~6~], and the possibility of
fractional statistics on the sphere [~7,~8~]. We consider two or three
spins coupled antiferromagnetically.
Although a {\it naive} \sc analysis reproduces the low-lying
energies and degeneracies correctly, there is a marked difference between the
ways in which the naive \sc and spin \wf transform under the
\ex of any two spins. The
complete symmetry group is the  permutation group \sn of $N$ spins in
the \qu theory, and the  braid group \bn on the sphere in the naive
\sc theory.
We relate this difference in transformation properties to a WZ term which
appears in the coherent state \pain \la for spins. This term treats each spin
as a charged particle moving on a sphere which has a magnetic monopole of
strength $4 \pi S~$ at its center. Once this term is taken into account, the
{\it modified} \sc \wf have the
correct transformation properties under spin permutations.
We can think about the \sc problem in terms of $N$ anyons moving on a
sphere with the statistics parameter $\theta~$ (which is defined modulo
$~2\pi~$) being $~2\pi S~$ for $~N~=~2$ and $~3 \pi S~$ for $~N~=~3$.
Thus the naive \sc \wf for three half-integer spins exhibit {\it semionic}
statistics. At the end, we will verify our analysis by directly constructing
the correct \sc \wf from the spin \wf.

Consider first an antiferromagnetic \ha for two spins
$~H ~=~ (\son + \stw )^2 ~\equiv ~\lo^2~$.
For any value of the spin $S$, the energies are $~l(l+1)~$ with a
degeneracy $~2l+1~$ where $~l~=~0,1,2,~...~,~2S~$. The three-$j$ symbols have
the symmetry [~9~]

\vfill
\eject

$$\left( \matrix{S &~S &~l \cr
m_1 &~m_2 &~m \cr} \right) ~~=~ (~-~1~)^{~2S~+~ l}~~ \left(
\matrix{S &~S &~l \cr
m_2 &~m_1 &~m \cr} \right) $$
where $~m_1~,~~m_2~$ and $~m~$ denote eigenvalues of $~(~\son ~)_3 ~, (~
\stw~)_3~$ and $~l_3~$ respectively.
Hence, under the \ex $~P_{12} ~\equiv~ \son ~\leftrightarrow ~\stw~$, the \wf
transform by the phase $~(~P_{12} ~)_{qu} ~=~ (~-1~)^{~2S~+~l}~$
where the subscript `qu' denotes \qu.

Now consider a \sc treatment of this problem. For $S >> 1~$,
we introduce a vector $~\ph ~=~ (~\son ~-~ \stw~)~/~2S ~$. This satisfies
$$\eqalign{\lo \cdot \ph ~&=~0 \cr
{\rm and}~~~ \ph^2 ~&=~ 1 ~+~ {1 \over S} ~-~ {{\lo^2} \over {4 ~S^2}} \cr}
\eqno(1)$$
For low-lying excitations (i.e. $~l~<<~S~$), we see from (1) that
$\ph$ is an unit vector.  The naive \sc \la is
$${\cal L} ~=~ {1 \over 4} ~{\dot \ph}^{~2}
\eqno(2)$$
with the constraint $~\ph^2 ~=~1~$. Canonical quantization of (2)
reproduces the above Hamiltonian except that $~- ~\lo^{~2} ~$ is now given
by the Laplacian $~{\vec \nabla}^2 ~$ on a sphere.
The \sc energies are therefore $l(l+1)$ with degeneracy $2l+1~$, and the
naive \wf are the spherical harmonics $Y_{l,m}~(\alpha,~\beta)~$.
(Here $~(\alpha,~\beta)~$ are the polar angles such that $~\ph ~=~ (\sin~
\alpha ~\cos ~\beta, ~\sin~\alpha ~\sin ~\beta, ~\cos~
\alpha ~)~$. Namely, the direction of spin (or particle) $1$
has the coordinates
$~(\alpha,~\beta)~$ while particle $2$ is at the point $~(\pi ~-~ \alpha,~
\pi ~+~\beta)~$ where $0~\le~ \alpha~\le~\pi, ~~0~\le~ \beta ~<~ 2 \pi~$).
Semiclassically,
$l$ can be any non-negative integer. This spectrum agrees with the exact one
for $~0~\le ~l~\le ~2S~$. It is interesting that the \sc energies and
degeneracies are correct even if $S$ is not much greater than one and $l$
is not much less than $S$. Under the \ex of the two spins, $\ph \rightarrow~
-~ \ph$. The  \wf $~Y_{l,m}~$ then transform by the phase
$(~P_{12}~)_{nsc} ~=~ (~-1~)^{~l}~$
where the subscript `nsc' denotes naive \sc.

The difference of $~(~-1~)^{~2S} ~$ between $~(~P_{12}~)_{qu}~$
and $~(~P_{12}~)_{nsc}~$ can be explained as
follows. The two-spin problem can be semiclassically thought of as two
particles moving on a sphere with the \ha forcing them
to lie at antipodal points for low energies. Under an \ex, the two
particles will together trace out a closed curve which encloses a solid
angle $2\pi$. Now, it is known that the path integral for
spins contains a WZ term which makes each particle see a magnetic
monopole of strength $~4\pi S~$ at the center of the sphere [~2,~6~]. A
particle which goes around a closed curve enclosing a solid angle
$\Omega$ picks up an Aharonov-Bohm phase $~\exp~(i\Omega S)~$. Thus an
\ex of the two particles produces a phase $~\exp~(i2\pi S)~$.

To be explicit, the WZ term equals $~S~{\dot \beta}~(1~+~{\rm cos} ~\alpha~)~$
for particle $1$ if we choose the Dirac string of the monopole to pass through
the north pole $~\alpha~=~ 0~$. The WZ term for particle $2$ is then
$~S~{\dot \beta}~(1~-~{\rm cos} ~\alpha~)~$. The sum of the two is
$${\cal L}_{WZ} ~=~ 2~S~{\dot \beta}
\eqno(3)$$
which is a total derivative. The correct \sc \wf are therefore
$\exp~(i2 \beta S)~ Y_{l,m}~(\alpha, ~\beta)~$. Note that since the
phase factor $~\eta~=~ \exp~(i2 \beta S)~$ is single-valued (except
at the points $~\alpha ~=~ 0 ~~{\rm or}~~ \pi~$ when one of the
particles lies on the Dirac string), the {\it energy spectrum} is
unaffected by (3). But the new \sc \wf do show the correct \ex phase
$~(P_{12}~)_{qu}~$ due to the factor $~\eta~$.

To summarize the \sc picture, the particles behave like anyons on a
sphere with the parameter $\theta ~=~2\pi S~$. For $N$ anyons on a
sphere, $\theta$ is only allowed to have the $2(N-1)$ values given by
$~\pi p/(N-1) ~$, where $p~=~0,1,2,~...~,~2N-3 ~$ [~8~]. So we may hope to
find a truly anyonic behavior (i.e. $\theta \ne 0 ~~{\rm or}~~ \pi~$)
if $~N~=~ 3~$. We therefore turn to the more interesting problem of
{\it three} spins.

The \ha $~H ~=~ (\son + \stw + \sth)^2 ~\equiv ~\lo^2 ~$
can be shown to have the low-lying spectrum $l(l+1)$ with degeneracy
$(2l+1)^2 ~$ for $~0~\le~ l~\le~ S~$. (For $~l~>~S ~$, the expression for
degeneracy is different). Here $l$ takes integer or half-integer values if $S$
is an integer or half-integer respectively.
For any $l$, let us consider the $2l+1$ states which have the
eigenvalue $l_3 ~=~l~$, and study their transformation properties under
the \pg $~S_3~$ of the three spins. (Since the group operations of $S_3~$
commute with the total spin operators $~\lo~$, the
same transformation properties
will hold for other values of $~l_3~$  also). We have only examined small
values of $S$ and $l$, namely, $~0~\le ~l~\le ~S~\le ~2~$. For integer $S$,
we find that these $2l+1$ states fall into $l+1$ irreducible representations
(IR) of $S_3~$ consisting of $l$ doublets and one singlet. Under the \ex of
any two spins, the singlet picks up the phase $~(~P~)_{qu} ~=~
(~-1~)^{~S~+~l}~$.
(For example, the ground state $(~l~=~0~)~$ has a totally symmetric wave
function if $S$ is an {\it even} integer, and an antisymmetric wave
function if $S$
is an {\it odd} integer). For half-integer $S$, the $2l+1$ states
with $l_3 ~=~ l~$ fall into
$~l+~1/2~$ doublets. Under any \ex, the doublets always transform by a
$~2~\times~2~$ matrix whose eigenvalues are $~\pm ~1~$.

Some of the above statements can be understood using the {\it two-spin}
results. To obtain a total spin $l$ with three spins, the spin of two of
them, say $\son ~$ and $\stw ~$, must add up to values lying in the
range $~S~-~l,~...~,~S~+~l ~$. (This explains the $(2l+1)$-fold
degeneracy for a given value of $l_3~$). Under the {\it subgroup} $P_{12}~$,
the \ex phases are therefore $~(~-1~)^{~3S ~-~ l}~,~...~,~(~-1~)^{~3S ~+~
l}~$. It then follows that for half-integer $S$, there are $~l~+~1/2~$ states
with $~P_{12}~ =~-~1~$ and $~l~+~1/2~$ with $~P_{12}~ =~1~$, while for
integer $S$, there are $l$ states with $~P_{12}~=~ (~-1~)^{~S~+~l~+~1}~$ and
$~l~+~ 1~$ with $~P_{12}~=~ (~-1~)^{~S~+~l}~$. Of course, it needs more work
to derive the singlet and doublet structure under the {\it full} group
$~S_3~$.

Now we do a \sc analysis. For large $S$, we introduce two vectors
[~4~]

\vfill
\eject

$$\eqalign{\phon ~&=~ {{\son~-~\stw} \over {{\sqrt 3} S}} \cr
{\rm and}~~~ \phtw ~&=~ {{\son + \stw - 2 \sth} \over {3 S}} \cr}
\eqno(4)$$
One can again derive identities similar to (1) which show that
for $~S~$ much greater than both $~l~$ and $~1~$,
$~\phon^2 ~=~ \phtw^2 ~=~1~$ and
$~\phon \cdot \phtw ~=~ 0~$. We introduce a third vector $\phth ~=~
\phon ~\times~ \phtw ~$ and define an $SO(3)$ matrix $~\romat~$ whose columns
are given by the three vectors $~\phon ~,~\phtw $ and $\phth~$
respectively. The total spin $~\lo~$ transforms $~\romat~$ from the {\it left}.
Thus $~[~l_i~, ~\romat ~] ~=~-~ T_i ~\romat~$
where $i~=~ 1,2,3 ~$ and the matrices $~T_i~$ have the elements
$~(~T_i~)_{jk} ~=~ -~i~{\epsilon_{ijk}} ~$.

There is a two-to-one mapping from the manifold of $SU(2)$, namely the
hypersphere $S^3~$, to the matrices $\romat~$. (This is made explicit in Eq.
(13) below).  We parametrize points on
$~S^3~$ by an $SU(2)$ matrix $~\vmat ~=~ x_4 ~+~ i~{\vec \sigma} \cdot {\vec
x}~$
where ${\vec \sigma}$ are the Pauli matrices and $~x_4^2 ~+~
x_1^2 ~+~x_2^2 ~+~x_3^2 ~=~1~$. For later use, we define
$$\eqalign{z_1 ~&=~ x_1 ~+~ i~x_2 ~=~\sin ~a ~e^{i b} \cr
{\rm and}~~~~ z_2 ~&=~ x_3 ~+~ i~x_4 ~=~\cos ~a ~e^{i c} \cr}
\eqno(5)$$
where $~0 ~\le ~a ~\le~ \pi /2~$ and $~0 ~\le~ b, ~c ~<~ 2
\pi ~$. The symmetry group of the manifold $S^3~$ is $~SO(4) ~=~ SU(2) ~\times~
SU(2)~/~Z_2 ~$. These two $SU(2)$'s are generated by matrices $~\lo~$ and
$~{\vec l}^{~\prime}~$ which act on $~\romat~$ (or $~\vmat~$)
from the left and right respectively [~10~]. (The two can be thought of as
rotations about a set of axes which is either space-fixed or body-fixed).
On $~\romat~$, a rotation acts as an orthogonal
matrix $~{\underline O} ~=~ \exp~(i \epsilon ~ {\hat n} \cdot {\vec T} ~)~$
while on $~\vmat~$, the corresponding action is by an unitary matrix
$~{\underline U} ~=~ \exp~(~i \epsilon ~ {\hat n} \cdot
{\vec \sigma}~/2 ~)~$ where $~{\hat n}~$ is an unit vector and $~\epsilon~$ is
the rotation angle. (Notice that we
are using the same symbol $~\lo~$ to denote both the \qu spin operators
and the \sc angular momenta which act from the left. It will be clear
from the context which one we mean).

The \sc \la can now be shown to be [~4,~11~]
$${\cal L} ~=~ {1 \over 8}~~ {\rm tr} ~{\dot \romat}^T ~{\dot \romat}~=~
{1 \over 2}~ {\rm tr}~{\dot \vmat}^{\dag} ~{\dot \vmat}
\eqno(6)$$
where $\romat^T ~=~ \romat^{-1} ~$ and $\vmat^{\dag} ~=~ \vmat^{-1} ~$.
Canonical quantization of this yields
$~H ~=~ -~ {\vec \nabla}^2 ~/~4 ~$
where $~{\vec \nabla}^2~$ is the Laplacian on  $~S^3~$. The
normalization in  Eq. (6) is fixed by considering small
fluctuations near the identity. $~\romat ~=~ I_3 ~+~ i~2~{\vec \epsilon}
\cdot {\vec T} ~$ corresponds to $\vmat ~=~ I_2 ~+~ i~{\vec \epsilon}
\cdot {\vec \sigma} ~$ which is near the north pole of $S^3~$,
namely, $x_4 ~=~ 1~$.
(Here $~I_n~$ denotes the $n ~\times~n~$ identity matrix). Then (6)
becomes ${\cal L} ~=~ {\dot {\vec \epsilon}}^{~2} ~$, the \ha is
${\vec \Pi}_{\epsilon}^{~2} ~/4~$, and the Laplacian on $S^3~$ is
${\vec \nabla}_{\epsilon}^{~2} ~=~ -~{\vec \Pi}_{\epsilon}^{~2} ~$.

On $~S^3~$, the left operators $~{\vec l}~$ have the form [~10~]
$$\eqalign{l_3 ~&=~{1 \over 2}~ (~z_1 ~\partial_1 ~+~ z_2 ~\partial_2 ~-~
z_1^\star ~{\partial_1^\star} ~-~ z_2^\star ~{\partial_2^\star} ~)~
= ~{1 \over {2i}} ~(~{\partial \over {\partial b}} ~+~{\partial \over
{\partial c}} ~) \cr
l_+ ~&=~ z_2~ {\partial_1^\star}~-~ z_1~ {\partial_2^\star} ~~~~{\rm
and}~~~~ l_- ~=~ z_1^\star~ \partial_2~-~ z_2^\star~ \partial_1 \cr}
\eqno(7a)$$
while the right operators $~{\vec l}^{~\prime}~$ are
$$\eqalign{{l_3}^{\prime} ~&=~{1 \over 2}~ (~z_1 ~\partial_1 ~-~
z_2 ~\partial_2 ~-~ z_1^\star ~{\partial_1^\star} ~+~ z_2^\star ~
{\partial_2^\star} ~)~ = ~{1 \over {2i}} ~(~{\partial \over {\partial
b}} ~-~{\partial \over {\partial c}} ~) \cr
{l_+}^{\prime} ~&=~ z_2^\star~ {\partial_1^\star}~-~ z_1~
\partial_2 ~~~~{\rm and}~~~~ {l_-}^{\prime} ~=~ z_1^\star~
\partial_2^\star ~-~ z_2 ~ \partial_1 \cr}
\eqno(7b)$$
where $\partial_i ~$ denotes $\partial ~/~\partial z_i ~$.
The Laplacian is $~{\vec \nabla}^2 ~=~-~4~{\vec l}^{~2} ~=~-~4~{\vec l}^{~
\prime 2} ~$.
The harmonic functions  on $S^3~$ transform as the representation
$(~l,~l~)~$ under $SU(2) ~\times SU(2)~$. Here $l$ can be an integer
or half-integer. The degeneracy is $(2l+1)^2~$. (For instance, the
first five harmonics are $1$ for $l~=~ 0~$, and $z_1 ~, ~z_2 ~, ~z_1^\star ~~
{\rm and}~~ z_2^\star ~$ for $l~=~ 1/2 ~$). The Laplacian acting on
these harmonics gives $~-~4l(l+1)~$. Under a $2\pi ~$ rotation of the
spins about any axis, $~\romat ~\rightarrow~
\romat ~, ~~~ \vmat ~\rightarrow ~-~
\vmat ~$ and the harmonics transform by $(~-1~)^{~2l} ~$. So we have to choose
$l$ to be an integer (or half-integer) if $S$ is an integer (or
half-integer). Thus the \sc theory reproduces the correct spectrum
for $0 ~\le ~l~\le ~S~$. The proper manifold for \sc quantization
is therefore $SO(3)$ for integer $S$ and $SU(2)$
for half-integer $S$ as we might
have expected for an odd number of spins.

The $2l+1$ \sc \wf with $l_3 ~=~ l~$ are $~z_1^{2l} ~,~z_1^{2l-1}~
z_2~,~ ...~,~z_2^{2l} ~$.  An \ex of
any two spins is equivalent to a $180^o~$
rotation about the third spin which is a matrix acting on $~\romat~$
or $~\vmat~$ from the {\it right}. (For $~\vmat~$, one has
to further specify whether the rotation is clockwise or anticlockwise as the
corresponding matrices differ by a minus sign).
Under any one of the three possible $180^o~$ rotations,
we find that
$$\left(\matrix{z_1 \cr
z_2 \cr}\right) ~~~\rightarrow~~~ {\underline P}~~\left(\matrix{z_1 \cr
z_2 \cr}\right)
\eqno(8)$$
where  ${\underline P}$ is an off-diagonal unitary matrix whose eigenvalues are
$~\pm ~i~$, {\it not} $~\pm 1~$. For example, an \ex of $\son$ and $\stw$
corresponds to multiplying $~\romat~$ from the right by
$$\left(\matrix{-1 &0 &0 \cr
0 &1 &0 \cr
0 &0 &-1 \cr}\right)$$
and $\vmat$ from the right by $~i~\sigma^2 ~$
or $~-~i~\sigma^2 ~$. So the matrix
in (8) is $~{\underline P}_{12} ~=~ i~\sigma_2 ~$ or $~ -~i ~\sigma_2 ~$.

Hence, for half-integer $l$, the $2l+1$ functions given above fall
into $~l+ 1/2~$ doublets each transforming irreducibly under the {\it braid}
group $B_3~$, namely,
$$\left(\matrix{z_1^{2l} \cr
z_2^{2l} \cr}\right) ~~,~~\left(\matrix{z_1^{2l-1}~z_2 \cr
z_1 ~z_2^{2l-1} \cr}\right) ~~,~~...
\eqno(9)$$
Under any \ex, these doublets transform with eigenvalues $~\pm ~i~$.
For integer $l$, on the other hand, the $2l+1$
states fall into $l$ doublets as in Eq. (9),
but there is also a singlet given by $~z_1^l ~z_2^l ~$. Under any \ex,
the doublets transform with eigenvalues $~\pm ~1~$ while the singlet
picks up the phase $~(~P~)_{nsc} ~=~(~-1~)^{~l}~$.
Thus in all cases, the eigenvalues of the (one- or two-dimensional) \ex
matrices $(~{\underline P}_{ij}~)_{nsc}~$ differ from those of
$~(~{\underline P}_{ij}~)_{qu}~$ by the phase $~\exp~(i3\pi S)~$.

We can understand this by using the WZ term and a typical \ex shown in
Fig. 1. The antiferromagnetic \ha forces the three particles to lie
$120^o ~$ apart on a sphere. Assume that $\sth$ is fixed along the south
pole $~(~0,~0,~-1~)~$ and \ex $\son$ and $\stw$ by a $180^o ~$
rotation. The resultant closed curve on the sphere encloses the solid angle
$~\Omega ~=~2 \pi ~(~1~+~ \cos~ {60^o} ~) ~=~ 3\pi~$. Hence the
Aharonov-Bohm phase picked up is $~\exp~(i3\pi S)~$. Note that the sense
of the \ex is important if $S$ is an half-integer. In this Letter, however,
we will not be precise about whether the argument of the phase
is $~3\pi S~$ or $~-~3\pi S~$.

The symmetry group of three particles placed $120^o ~$ apart on the sphere
is the \bg $B_3~$. A one- or two-dimensional IR of $B_3~$
can be obtained by taking an IR of the \pg $S_3~$ and making it anyonic
by multiplying the appropriate matrices by $\exp~(i \theta )~$ for a
clockwise \ex of any two particles. What we have shown above is that
the anyon parameter $\theta$ is equal to $~\pm  ~3 \pi S ~$.

We now study how these naive \wf get modified if the WZ term is included in
the \la. We first parametrize the three classical spins ${\vec S_n}~$ in terms
of the orthonormal vectors $~{\vec \phi}_n ~$ given by
$$\eqalign{\phth ~&=~(~\sin~ \alpha ~\cos~ \beta, ~\sin~\alpha ~\sin~ \beta,~
\cos~ \alpha ~) \cr
\phon ~&=~(~-~\cos~ \alpha~ \cos~ \beta, ~-~\cos~ \alpha ~\sin~ \beta,~ \sin~
\alpha ~) \cr
{\rm and}~~~~ \phtw ~&=~(~\sin~ \beta, ~-~\cos~ \beta,~0 ~) \cr}
\eqno(10)$$
Then
$${\vec S_n} ~/~S ~=~ \sin ~(~ \gamma ~+~ {{2\pi n} \over 3}~)~ \phon ~-~
\cos ~(~ \gamma~+~ {{2\pi n} \over 3}~)~ \phtw
\eqno(11)$$
for $~n~=~1,~2,~3~$. Now we find that the WZ term obtained by adding up the
contributions from the three spins is {\it again} a total derivative. Indeed,
$${\cal L}_{WZ} ~=~ S~ {d \over {dt}} ~\Bigl[~3~ \beta~-~3~ \gamma ~-~i~
\ln~\Bigl(~ {{\tan^3 ~(\alpha /2) ~-~i~\exp~(i3 \gamma )} \over {\tan^3 ~(
\alpha /2) ~+~i~ \exp~(-~i3 \gamma ~)}} ~\Bigl) ~\Bigr]
\eqno(12)$$

Next, we rewrite the matrix $~\romat~$
in terms of the $~S^3~$ coordinates (5). This takes the form
$$\romat ~=~ \left(~ \matrix{1~-~2~(~x_2^2~+~ x_3^2~) & ~~2~(~x_1 x_2~+~
x_3 x_4~) & ~~2~(~x_3 x_1 ~-~ x_2 x_4 ~) \cr
2~(~x_1 x_2 ~-~ x_3 x_4 ~) & ~~1~-~2~(~x_3^2~+~ x_1^2~) & ~~
2~(~x_2 x_3 ~+~ x_1 x_4 ~) \cr
2~(~x_3 x_1 ~+~ x_2 x_4 ~) & ~~2~(~x_2 x_3 ~-~ x_1 x_4 ~) & ~~1~-~2~(~x_1^2~+~
x_2^2~) \cr} ~\right)
\eqno(13)$$
A comparison then shows that $~ \alpha ~=~2~ a, ~~~\beta ~=~ b ~+~ c~~~~
{\rm and}~~~~\gamma ~=~ c~-~ b ~$. The WZ term is thus
$${\cal L}_{WZ} ~=~-~i~S~ {d \over {dt}} \ln \Bigl( ~{{z_1^3 ~-~i~ z_2^3}
\over {{z_1^\star}^3 ~+~i~ {z_2^\star}^3} } ~\Bigr)
\eqno(14)$$
Hence the correct \sc \wf are obtained by multiplying the naive \wf by the
phase factor
$$\eta ~=~ \Bigl( ~{{z_1^3 ~-~i~ z_2^3} \over {{z_1^\star}^3 ~+~i~
{z_2^\star}^3} } ~\Bigr)^S
\eqno(15)$$
(Our choice for the direction of the Dirac string and  the expression
for $~\eta~$ seem to break rotational invariance. However, one can define
new angular momenta by the unitary transformation $~{\vec l}~
\rightarrow~ \eta~ {\vec l}~ \eta^{~-1} ~$).

We see that $~\eta~$ is single-valued on $S^3~$ (unless $~z_1^3 ~=~ i~
z_2^3 ~$ which corresponds to one of the particles being at the north pole
of the sphere). Hence the spectrum remains unchanged. But the phase factor
restores the correct permutation properties to the \wf. For instance, under the
exchange $~P_{12}~$,  $~z_1 ~\rightarrow  ~z_2~$ and  $~z_2 ~\rightarrow ~-~
z_1~$. Then $~\eta~$ changes by $~\exp~ (~- ~i 3 \pi S)~$ as required.
(See the discussion following Eq. (21) below. If $~P_{12}~$ is taken to
transform $~z_1 ~\rightarrow  ~-~z_2~$ and $~z_2 ~\rightarrow ~
z_1~$, then $~\eta~$ changes by $~\exp~ (i 3 \pi S)~$).

Finally, we turn to a derivation of the \sc \wf from the spin \wf
for $~S= 1/2~$. For the
direction $~{\hat n} =~ (~\sin~ \alpha ~\cos~ \beta, ~\sin~ \alpha ~\sin~
\beta,~ \cos~\alpha ~)$, the
eigenvector of $~{\hat n} ~\cdot ~{\vec \sigma}~/2 ~$ with eigenvalue $~1/2~$
is given by
$$\vert ~\alpha,~\beta~ \rangle ~=~ \left(~ \matrix{\cos ~(\alpha/2) ~~exp~(-~
i \beta) \cr
\sin ~(\alpha/2) \cr} ~\right)
\eqno(16)$$
The phase is chosen such that (16) is ill-defined only at the north pole.
The {\it ket} $~\vert ~ \ph ~ \rangle ~$ for the two-spin problem is defined
to be the tensor product
$$\vert ~ \ph ~ \rangle ~\equiv ~\vert ~\alpha,~\beta~ \rangle ~~ \otimes ~~
\vert ~ \pi ~-~ \alpha,~ \pi ~+~\beta~ \rangle
\eqno(17)$$
where the first and second factors denote the vectors (16) for spins $1$ and
$2$ respectively. The exact ground state with $~l~=~l_3~=~0~$ has the wave
function
$$\vert ~\psi_{0,0} ~\rangle ~=~ {1 \over {\sqrt 2}} ~\left( \matrix{1 \cr
0 \cr} \right) ~\otimes ~\left( \matrix{0 \cr
1 \cr} \right) ~~-~~ {1 \over {\sqrt 2}} ~\left( \matrix{0 \cr
1 \cr} \right) ~\otimes ~\left( \matrix{1 \cr
0 \cr} \right)
$$
We then {\it define} the \sc wave function to be the amplitude
$\langle ~\ph ~\vert ~\psi_{0,0} ~ \rangle~$ which turns out to be
$~\exp~(i \beta)~  Y_{0,0} ~(\alpha, ~ \beta) ~/~ {\sqrt 2}~$.
Similarly, the three excited states with $~l~=~1~$  have the \wf
$\langle ~\ph ~\vert ~\psi_{1,m} ~ \rangle~=~\exp~(i \beta) ~Y_{1,m} ~/~
{\sqrt 6}~$.
(The $~Y_{l,m} ~$ are normalised according to the measure $~d\ph ~=~ \sin~
\alpha~~ d \alpha ~d \beta~/~4 \pi ~$).

For the three-spin problem, we define the ket $~\vert ~\vmat~ \rangle~$ as the
tensor product
$$\vert ~\vmat~ \rangle~=~ \vert ~\alpha_1~,~ \beta_1 ~\rangle ~~\otimes~~
\vert ~\alpha_2~,~ \beta_2 ~\rangle ~~\otimes ~~\vert ~\alpha_3~,~
\beta_3 ~\rangle
\eqno(18)$$
where the three kets $~\vert ~\alpha_n~, \beta_n ~\rangle~$ can be deduced
from Eq. (11). To be explicit, the {\it bra} $~\langle ~\alpha_n~, \beta_n ~
\vert ~$ takes the form
$$\langle ~\alpha_n~, \beta_n ~\vert ~=~{\sqrt {{z_1 ~+~ i \omega^n ~z_2}
\over {2~(~z_1^\star ~-~ i \omega^{2n} ~z_2^\star)}} } ~~\Bigl( ~z_1 ~+~ i
\omega^{2n} ~z_2 ~~~~~~~z_1^\star ~-~ i \omega^{2n} ~z_2^\star ~\Bigr)
\eqno(19)$$
where $~\omega ~=~ \exp~(i 2 \pi/3)~$.
It is then clear that any $~3$-spin wave function $~\vert~ \psi~ \rangle~$
will have an amplitude $~\langle ~\vmat~ \vert~\psi ~\rangle ~$ which is a
polynomial in $~z_i ~$ and $~z_i^\star ~$ multiplied by the phase $~\eta~$ in
Eq. (15). For instance, consider the two ground states
with $~l~=~l_3~= ~1/2 ~$. Their \wf
$$\eqalign{\vert ~\psi_1 ~\rangle ~=~& {1 \over {\sqrt 2}} ~\left(
\matrix{1 \cr 0 \cr} \right) ~\otimes ~\left( \matrix{0 \cr
1 \cr} \right) ~\otimes ~\left( \matrix{1 \cr
0 \cr} \right) ~-~ {1 \over {\sqrt 2}} ~\left( \matrix{0 \cr
1 \cr} \right) ~\otimes ~\left( \matrix{1 \cr
0 \cr} \right) ~\otimes ~\left( \matrix{1 \cr
0 \cr} \right) ~~~~{\rm and} \cr
\vert ~\psi_2 ~\rangle ~=~& {1 \over {\sqrt 6}} ~\left( \matrix{1 \cr
0 \cr} \right) ~\otimes ~\left( \matrix{0 \cr
1 \cr} \right) ~\otimes ~\left( \matrix{1 \cr
0 \cr} \right) ~+~ {1 \over {\sqrt 6}} ~\left( \matrix{0 \cr
1 \cr} \right) ~\otimes ~\left( \matrix{1 \cr
0 \cr} \right) ~\otimes ~\left( \matrix{1 \cr
0 \cr} \right) \cr
&-~ {2 \over {\sqrt 6}} ~\left( \matrix{1 \cr
0 \cr} \right) ~\otimes ~\left( \matrix{1 \cr
0 \cr} \right) ~\otimes ~\left( \matrix{0 \cr
1 \cr} \right) \cr}
\eqno(20)$$
transform under the \ex $~P_{12}~$ with phases $~-~1$ and $~1~$
respectively. One then finds that
$$\eqalign{\langle ~\vmat~ \vert ~\psi_1 ~\rangle ~=~ {{\sqrt 3} \over 4} ~(~
z_2 ~+~ i~z_1~)~~ \eta \cr
{\rm and} ~~~~\langle ~\vmat~ \vert ~\psi_2 ~\rangle ~=~ {{\sqrt 3} \over 4} ~
(~z_1 ~+~ i~z_2~)~~ \eta \cr}
\eqno(21)$$
If $~P_{12}~$ takes $~z_1 ~\rightarrow ~z_2 ~$ and $~z_2 ~\rightarrow ~-~
z_1 ~$, then $~\eta ~\rightarrow ~i~\eta ~$ (that is, $~\eta~ \rightarrow ~
\eta ~\exp~(-~i3 \pi S)~$). Hence $~\langle ~\vmat~\vert ~\psi_1 ~
\rangle ~$ and $~\langle ~\vmat~\vert ~
\psi_2 ~\rangle ~$ transform correctly under $~P_{12} ~$.

This construction can be generalized to any spin $~S~$. Let us work in a
basis in which $~(~{\vec S}_n ~)_3 ~$ is diagonal. The key ingredient is the
eigenvector $~\vert~ \alpha, ~\beta ~\rangle~$ (with eigenvalue $~S~$)
of the
matrix $~{\hat n} \cdot {\vec S}~$. The $~m^{\rm th}~$ entry of the column
$~\vert~ \alpha, ~\beta ~\rangle~$ is given by [~9~]
$$\vert~ \alpha, ~\beta ~\rangle_{m} ~=~ {\sqrt {{2~S} \choose {S ~+~ m}}}~~
\Bigl(~\cos ~{\alpha \over 2} ~\Bigr)^{~S ~+~ m} ~\Bigl(~\sin ~{\alpha
\over 2} ~\Bigr)^{~S ~-~ m} ~e^{-~i(S+m) \beta}
\eqno(22)$$
where $~m~$ takes the values $~S, ~S-1, ~...~, ~-~S~$ from top to bottom.
This is well-defined everywhere except at the north pole. Then the two-spin
ket in (17) produces \wf of the form
$$\langle~ \ph ~\vert ~\psi_{l,m}~ \rangle ~=~ N_{S,l,m} ~~
e^{i2 S \beta} ~~Y_{l,m}
\eqno(23)$$
where $~N_{S,l,m}~$ is a normalization constant. To prove (23), we observe
that in terms of the Euler angles $~(\alpha, ~\beta, ~\gamma)~$, the vectors
$~\exp~(i S \beta)~ \vert ~\alpha, ~\beta ~\rangle ~$ and $~\exp~(i S \beta)
\vert ~\pi ~-~ \alpha, \pi ~+~ \beta ~\rangle ~$ are given by
the first and last columns of the rotation matrix $~{\cal D}^{(S)} ~(\alpha, ~
\beta,~\gamma)~$ respectively. These two columns are proportional to $~\exp~(-
i S \gamma)~$ and $~\exp~(i S \gamma)~$. The $~\gamma$-dependence cancels
when we take a direct product of the two. Thus $~\exp~(i 2 S \beta) ~\vert ~
\ph ~\rangle ~$ is given by a $~(2S+1)^2 ~$-dimensional column of the direct
product $~~{\cal D}^{(S)} ~(\alpha, ~\beta,~0) ~~\otimes~$

\noindent
${\cal D}^{(S)} ~(\alpha, ~\beta,~0) ~$. The transformation properties
of this object under rotations and consequently
the {\it differential equation} satisfied by it are then precisely the same
as those of the spherical harmonics $~Y_{l,m} ~(\alpha, ~\beta) ~$. (If
there had been a dependence on $~\gamma~$, then the transformations and the
differential equation would have been different since they would
involve $~\partial / \partial \gamma ~$  [~9~] ~). Thus,
$$e^{-~i2S \beta}~~ \langle~ \ph ~\vert ~\lo ~\vert ~\psi_{l,m} ~
\rangle ~=~ \lo ~~e^{-~i2S \beta}~~ \langle~ \ph ~\vert ~ \psi_{l,m} ~\rangle
\eqno(24)$$
On the left hand side of (24), $~\lo ~=~ \son ~+~ \stw~$ is a matrix which
acts on $~\vert ~\psi_{l,m}~ \rangle~$ while on the right hand side, $~\lo~$
is a differential operator acting on $~\exp~(-~i2S \beta)~
\langle~ \ph ~\vert ~\psi_{l,m} ~\rangle~$. (For example,
$~l_3 ~=~ -~i~ \partial/ \partial \beta ~$). Then Eq. (23) follows from the
fact that $~\vert ~\psi_{l,m}~ \rangle~$ has $~\lo^2 ~=~ l(l+1)~$
and $~l_3 ~=~m ~$.

The constants $~N_{S,l,m} ~\equiv~ N_{S,l} ~$ in (23) are independent
of $~m~$ due to rotational symmetry. We can choose $~N_{S,l}~$
to be real and positive. The completeness of the $~\vert ~\psi_{l,m}~
\rangle~$ and the fact that $~\langle ~\ph ~\vert ~\ph~ \rangle~ =~1~$
means that
$$\sum_{l=0}^{2S} ~~(2l~+~1)~ N_{S,l}^{2} ~=~ 1 $$
Note that the $~(2S+1)^2 \times (2S+1)^2~$  matrix
$$M ~=~ \int ~d \ph ~~\vert ~\ph~ \rangle ~\langle ~\ph ~\vert$$
is {\it not} proportional to the identity. The orthonormality of the
$~\vert ~\psi_{l,m}~ \rangle~$ implies that the eigenvalues of $~M~$ are given
by the constants $~N_{S,l}^2 ~$. An explicit formula for $~N_{S,l} ~$ can be
derived by considering the state with $~l_3 ~=~ l~$. The corresponding
three-$j$ symbols are given by
$$\left( \matrix{S &S &l \cr
m &l-m &-l \cr} \right) ~=~ (-~1)^{~S+l-m} ~~\biggl[ ~\left(
\matrix{2l \cr
l \cr} \right) ~{{(2S-l) ! ~(S+m) ! ~(S+l-m) !} \over {(2S+l+1) ! ~(
S-m) ! ~(S-l+m) !}} ~\biggr]^{1/2}
$$
Using this one finds that
$$e^{-~i2S \beta} ~~\langle ~\ph ~\vert ~\psi_{l,l} ~\rangle ~\sim~
Y_{l,l}~ (\alpha, ~\beta) $$
and the constant of proportionality determines
$$N_{S,l}^2 ~=~ {{(2S) ! ~(2S) !} \over {(2S+l+1) ! ~(2S-l) !}}
\eqno(25)$$

For the ~three-spin problem, the ~ket in ~Eq. (18) produces
the ~\wf

\noindent
$\langle ~\vmat ~
\vert ~\psi_{l,m} ~\rangle$ which are given by polynomials in $~z_i ~$ and $~
z_i^\star ~$ times the phase in (15). From (16), (19) and (22), it is
clear that
the polynomial is even for integer $~S~$ and odd for half-integer $~S~$.
An analysis similar to the one for the two-spin case will show that
the polynomials are now given by
harmonic functions on $~S^3 ~$. We will however omit the details here.

It is worth noting that the \sc analysis in this Letter holds
for values of spin as small as $~1/2~$. It would be interesting to extend
these considerations to more than three spins [~5~] or even to a
spin chain. We often know, by either numerical or exact
methods, the way in which the quantum ground state and low-lying excitations of
antiferromagnetic spin chains transform under discrete symmetries like parity
(defined as the reflection of the chain about one site). One might ask how
this is  related to the symmetry properties of the
semiclassical field theories which are typically some non-linear sigma models.

\vskip .2in
\line{\bf Acknowledgments \hfill}

\vskip .1in

I thank Sumathi Rao for numerous discussions.

\vfill
\eject

\line{\bf References \hfill}
\vskip .3in

\item{1.}{F. D. M. Haldane, {\it Phys. Rev. Lett.} {\bf 50}, 1153 (1983)
and {\it Phys. Lett.} {\bf 93 A}, 464 (1983).}

\item{2.}{E. Fradkin, {\it Field Theories of Condensed Matter
Systems} (Addison-Wesley, New York, 1991)}.

\item{3.}{I. Affleck, in {\it Fields, Strings and Critical Phenomena},
Les Houches, 1988, ed. E. Brezin and J. Zinn-Justin (North Holland,
Amsterdam, 1990).}

\item{4.}{T. Dombre and N. Read, {\it Phys. Rev.} {\bf B 39}, 6797 (1989);~
S. Rao and D. Sen, IISc preprint IISc-CTS-92-9 (unpublished).}

\item{5.}{D. Loss, D. P. DiVincenzo and G. Grinstein, {\it Phys. Rev. Lett.}
{\bf 69}, 3232 (1992); J. von Delft and C. L. Henley, {\it Phys. Rev. Lett.}
{\bf 69}, 3236 (1992).}

\item{6.}{F. D. M. Haldane, {\it Phys. Rev. Lett.} {\bf 61}, 1029 (1988)
and {\it Phys. Rev. Lett.} {\bf 57}, 1488 (1986);~
T. Dombre and N. Read, {\it Phys. Rev.} {\bf B 38}, 7181 (1988);~
E. Fradkin and M. Stone, {\it Phys. Rev.} {\bf B 38}, 7215 (1988).}

\item{7.}{F. Wilczek, {\it Fractional Statistics and Anyon
Superconductivity} (World Scientific, Singapore, 1990).}

\item{8.}{T. D. Imbo and J. March-Russell, {\it Phys. Lett.} {\bf 252 B},
84 (1990); ~T. Einarsson, {\it Phys. Rev. Lett.} {\bf 64}, 1995 (1990).}

\item{9.}{E. P. Wigner, {\it Group Theory} (Academic Press, New York,
1959); ~J. D. Talman, {\it Special Functions} (W. A. Benjamin, New
York, 1968).}

\item{10.}{D. Sen, {\it J. Math. Phys.} {\bf 27}, 472 (1986); ~R. E.
Cutkosky, {\it J. Math. Phys.} {\bf 25}, 939 (1984).}

\item{11.}{A. P. Balachandran, G. Marmo, B.-S. Skagerstam and A. Stern,
{\it Classical Topology and Quantum States} (World Scientific,
Singapore, 1991).}

\vfill
\eject

\line{\bf Figure Caption \hfill}

\vskip .3in

\item{\bf 1.}{Three particles placed $120^o~$ apart on a sphere.
Exchanging particles $1$ and $2$ keeping particle $3$ fixed traces out a
closed curve which encloses the shaded area. The complement of this area
covers the solid angle $~3 \pi~$.}


\end